\documentclass{article}
\usepackage[english]{babel}
\usepackage[utf8]{inputenc}
\usepackage{geometry,amsmath,amssymb,graphicx}
\usepackage{listings}
\usepackage{verbatimbox}
\usepackage{caption}
\usepackage{subcaption}
\usepackage{xcolor}

\usepackage{amsthm}
\newtheorem{theorem}{Theorem}[section]
\newtheorem{definition}[theorem]{Definition}

\begin{document}

\title{Constant Delay Lattice Train Schedules}

\author{
    Jean-Lou De Carufel\thanks{School of Computer Science and Electrical Engineering, University of Ottawa, Ottawa, Canada} $^\ddag$ \and
    Darryl Hill\thanks{School of Computer Science, Carleton University, Ottawa, Canada} \and
    Anil Maheshwari$^\dag$\thanks{Research supported in part by NSERC} \and
    Sasanka Roy\thanks{Indian Statistical Institute, Kolkata, India} \and
    Luís Fernando Schultz Xavier da Silveira$^\dag$ 
}

\maketitle

\begin{abstract}
  The following geometric vehicle scheduling problem has been considered: given
  continuous curves $f_1, \ldots, f_n : \mathbb{R} \rightarrow \mathbb{R}^2$,
  find non-negative delays $t_1, \ldots, t_n$ minimizing $\max \{ t_1, \ldots,
  t_n \}$ such that, for every distinct $i$ {and $j$} and every time
  $t$, $| f_j (t - t_j) - f_i (t - t_i) | > \ell$, where~$\ell$ is a given
  safety distance.
  
  We study a variant of this problem where we consider trains (rods) of fixed length
  $\ell$ that move at constant speed and sets of train lines (tracks), each of which
  consisting of an axis-parallel line-segment with endpoints in the integer
  lattice $\mathbb{Z}^d$ and of a direction of movement (towards $\infty$ {or
  $- \infty$}). We are interested in upper bounds on the maximum delay we need
  to introduce on any line to avoid collisions, but more specifically on
  universal upper bounds that apply no matter the set of train lines.
  
  We show small universal constant upper bounds for $d = 2$ and any given
  $\ell$ and also for $d = 3$ and $\ell = 1$. Through clique searching, we are also able to show that several of these upper bounds are tight.
\end{abstract}

\section{Introduction}

Ajay et al. {\cite{AJAY2021}} considered the following problem: given a
set $S = \{ f_1, \ldots, f_n \}$ of curves, find a maximum subset $S'
\subseteq S$ such that, for any two distinct $f_i, f_j \in S'$, there is no
time~${0\leqslant{}t\leqslant{}1}$ when $f_i (t) = f_j (t)$. Though there are
some approximation algorithms, it is shown that most versions of these
problems are {{\textsf{NP}}-hard}. This is true even when the curves are
same-sized {{\textsf{L}}-shapes} and trajectories have the same constant
speed. During a talk by Sasanka Roy on this problem at Stony Brook University,
Joseph S. B. Mitchell and {Esther M. Arkin} posed the following variation:
given~$S = \{ f_1, \ldots, f_n \}$ where $f_i : \mathbb{R} \rightarrow
\mathbb{R}^2$ is a continuous function, we wish to assign delays $t_i
\geqslant 0$ so that, for any two distinct $f_i, f_j \in S$, there is no time
$t \in \mathbb{R}$ when $| f_j (t - t_j) - f_i (t - t_i) | \leqslant \ell$; we
also wish that the maximum delay is minimized.

\begin{figure}
\centering
	\includegraphics[page = 4, width = 7cm]{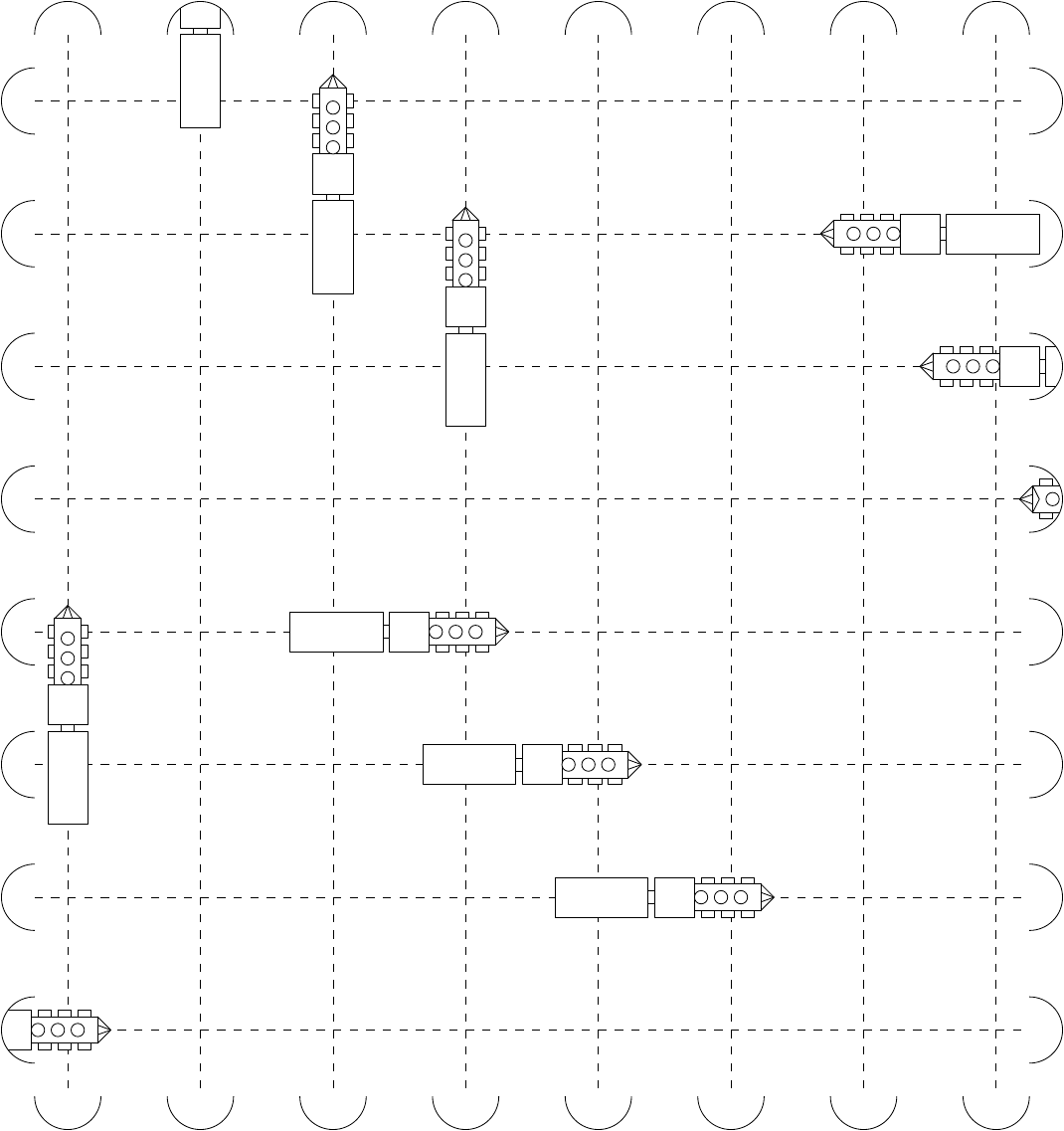}\caption{Train 
	lines on axis parallel tracks with stations (here seen as tunnels) 
	on the integer lattice.}\label{description}
\end{figure}

Automated guided vehicles (AGVs) and automated vehicle routing are currently
some of the most prolific fields in the domain of motor vehicle technology,
enjoying a large body of research addressing a large number of related
algorithmic and optimization problems. The field of automated guided vehicles
has a long and rich history and is closely related to our work. For an
elaborate review, the reader is referred to a survey by {Vis
{\cite{Vis2006}}}, though {\cite{HS2017}} is a more recent one. A central
topic in this field is collision avoidance. Kim and {Tanchoco {\cite{KT1991}}}
propose an algorithm based on Dijkstra's that facilitates conflict-free
routing of AGVs. It runs in $\mathcal{O} (v^4 n^2)$ time, where $v$ is the number
of vehicles and~$n$ is the number of nodes, i.e., path intersections. Arora
{et al. {\cite{ARM2000}}} used techniques based on game theory to design a
methodology for AGV traffic control. Yan {et al. {\cite{YZQ2017}}} studied
collision-free routing of AGVs on both unidirectional and bidirectional paths
using digraphs. For more information on conflict-free routing of AGVs, the
reader is referred to {\cite{Koff1987,Malmborg1990,ZWJ1991}}.
For an elaborate survey on routing and scheduling algorithms for AGVs, see
{\cite{QHHW2002}}. Other related work in this area includes the study of One
Way Road Networks (OWRNs) by Ajaykumar et al. {\cite{ADK2016}}, its
{motivation {\cite{Robbins1939}}} and
{\cite{KTM1978,DM2015,Scheffer2020}}.

\subsection{New Results}
In this paper, we study a simple vehicle scheduling problem in which, many times, we are guaranteed schedules with at most a constant delay.  We consider trains of fixed length $\ell$ that move at constant speed and sets of train lines (tracks), each of which consisting of an axis-parallel line-segment with endpoints in the integer lattice $\mathbb{Z}^d$ and of a direction of movement (towards $\infty$ {or $- \infty$}). For an example see Figure~\ref{description}. We are interested in upper bounds on the maximum delay we need to introduce on any line to avoid collisions, but more specifically on universal upper bounds that apply no matter the set of train lines.
  
We show small universal constant upper bounds for $d = 2$ and any given $\ell$ (Theorem \ref{thm:2d})  and also for $d = 3$ and $\ell = 1$ (Theorem \ref{thm:3d}). Through clique searching, we are also able to show that several of these upper bounds are tight (Section \ref{sec:clique}). In Appendix, we provide a Python code that we have used for clique searching, and the results are summarized in Table \ref{tab:lower-bounds}. 

\section{Preliminaries}

\begin{definition}
  A ($d$-dimensional) {{train line}} consists of:
  \begin{itemize}
    \item A {{track}}, which is a line segment in $\mathbb{R}^d$ with
    distinct endpoints distinguished between a departure point and an arrival
    point;
    
    \item A {{train length}}, which is a positive real number; and
    
    \item A {{speed}}, which is also a positive real number.
  \end{itemize}
  Furthermore, a set of train lines with non-overlapping (but possibly
  crossing) tracks is called a {{train network}}.
\end{definition}

\begin{definition}
  For a set of real numbers $X$ and a real number $y$, we denote $\{ x + y : x
  \in X \}$ by $X + y$.
\end{definition}
\begin{figure}[h]
    \centering
  \raisebox{-0.0388510421962379\height}{\includegraphics[width=7cm, page = 5]{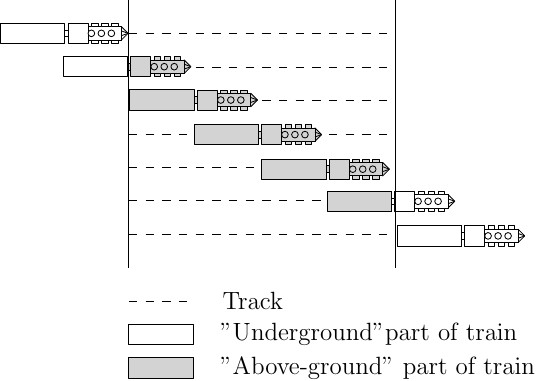}}
  \caption{\label{fig:train-journey}An illustration of the journey of a
  train.}
\end{figure}

\begin{definition}
  \label{def:schedule}A (collision-free) {{schedule}} for a train
  network is an assignment of a {{delay}}, which is a non-negative real
  number, to each of its lines. It also must result in no collisions between
  the trains. More precisely, for any two lines whose tracks cross, the open
  intervals $(0 ; \ell / v) + t + \delta / v$ and~$(0 ; \ell' / v') + t' +
  \delta' / v'$ must not intersect, where:
  \begin{itemize}
    \item $t$ and $t'$ are the respective delays assigned to the lines;
    
    \item $\delta$ and $\delta'$ are the respective distances between the
    departure points of the lines and the crossing point;
    
    \item $\ell$ and $\ell'$ are the respective train lengths of the lines;
    and
    
    \item $v$ and $v'$ are the respective speeds of the lines.
  \end{itemize}
  The {{delay}} of the schedule is the maximum delay it assigns. An
  {{integer schedule}} is one that only assigns integer delays.
\end{definition}

\begin{figure}[h]
\centering
  \raisebox{-0.585662707958562\height}{\includegraphics[width=7cm, page = 2]{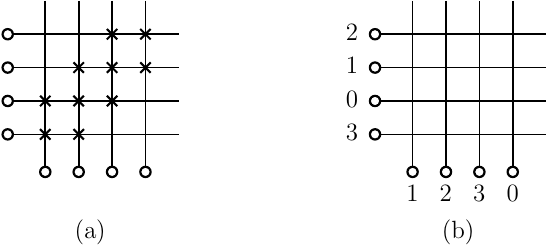}}
  \caption{\label{fig:example-schedule}If unit speed trains of length 2 depart
  simultaneously from the circles in (a), there will be collisions at the
  marked crossings. However, if delays are introduced as in (b), no collisions
  will occur.}
\end{figure}

An alternative way to understand schedules is to imagine each line's train
starts ``underground'' with the front at the departure point and moves towards
the arrival point at the line's speed, where it moves ``underground'' again
(``underground'' trains cannot collide with other trains); see Figure~\ref{fig:train-journey}. A delay assignment is then a schedule if, and only
if, the line segments that correspond to the ``above-ground'' parts of these
trains never cross. For an example, see Figure \ref{fig:example-schedule}.

\begin{definition}
  A $d$-dimensional train network is {{regular}} if all its lines'
  speeds are $1$, all its train lengths are the same and integer, all its tracks
  are axis-parallel and all its departure and arrival points are in
  $\mathbb{Z}^d$.
\end{definition}

\begin{theorem}
  \label{thm:regular-integer-schedule}All regular train networks admit an
  integer schedule of minimum delay (among all schedules, integer or
  otherwise).
\end{theorem}

\begin{proof}
  It is enough to show that, for an arbitrary schedule, assigning to each line
  a delay equal to the floor of the original delay will not result in any
  collisions. Indeed, take two lines whose tracks cross and define $t$, $t'$,
  $\delta$, $\delta'$, $v$, $v'$, $\ell$ and $\ell'$ as in Definition
  \ref{def:schedule}. We thus have $v = v' = 1$ and~$\ell = \ell' \in
  \mathbb{N}\backslash \{ 0 \}$. The original schedule ensures that
  \[ \bigl( (0 ; \ell) + t + \delta\bigr) \cap \bigl( (0 ;
     \ell) + t' + \delta' \bigr) = \{
     \}, \]
  which is equivalent to
  \[ (0 ; \ell) \cap \bigl( (0 ; \ell) + t' - t + \delta' -
     \delta\bigr) = \{ \} . \]
  Because $t' - t$ and $\lfloor t' \rfloor - \lfloor t \rfloor$ differ by less
  than $1$, $\lfloor t' \rfloor - \lfloor t \rfloor$ is {either $\lfloor t' -
  t \rfloor$} {or $\lceil t' - t \rceil$}. Therefore, since $\ell$ and
  $\delta' - \delta$ are integers, the open {interval $(0 ; \ell) + \lfloor t'
  \rfloor - \lfloor t \rfloor + \delta' - \delta$} cannot intersect $(0 ;
  \ell)$, which implies our delay assignment is collision-free.
\end{proof}

\begin{definition}
  Consider a train line with an axis-parallel track that departs {from $p \in
  \mathbb{R}^d$} and arrives at $p + \alpha e_i$, where $e_i$ is the $i$-th
  vector of the canonical basis of $\mathbb{R}^d$ ($0 \leqslant i < d$) and
  $\alpha \in \mathbb{R}\backslash \{ 0 \}$. The {{axis}} of the line
  is the number $i$. Furthermore, we say this line is {{positive}} if~$\alpha > 0$ and {{negative}} if~$\alpha < 0$, with its
  {{sign}} being $1$ or $- 1$, respectively.
\end{definition}

\begin{definition}
  For any modulus $k \in \mathbb{N}\backslash \{ 0 \}$, we extend the modulo
  $k$ function to real numbers as follows: {if $x \in \mathbb{R}$}, then
  $x\ensuremath{\operatorname{mod}}k = (\lfloor x \rfloor
  \ensuremath{\operatorname{mod}}k) + x - \lfloor x \rfloor$. Furthermore, for
  a set $X \subseteq \mathbb{R}$, we denote~$\{
  x\ensuremath{\operatorname{mod}}k : x \in X \}$ by
  $X\ensuremath{\operatorname{mod}}k$.
\end{definition}

\section{Results}

\begin{theorem}
  \label{thm:positive-lines}Any regular $d$-dimensional train network with
  trains of length $\ell$ and only positive lines has a schedule with delay at
  most $d \ell - 1$.
\end{theorem}

\begin{proof}
  Assign delay $\left( \ell a + \sum_{i = 0}^{d - 1} p_i \right)
  \ensuremath{\operatorname{mod}} (d \ell)$ to lines departing {from $p =
  (p_0, \ldots, p_{d - 1}) \in \mathbb{Z}^d$} with axis $a \in \{ 0, \ldots, d
  - 1 \}$. Consider then a crossing between a line departing from ${p = (p_0,
  \ldots, p_{d - 1}) \in \mathbb{Z}^d}$ with axis $a$ and another line
  departing from $p' = (p'_0, \ldots, p'_{d - 1}) \in \mathbb{Z}^d$ with axis
  $a' \neq a$. In the notation of Definition \ref{def:schedule}, we have:
  \begin{itemize}
    \item $t = \left( \ell a + \sum_{i = 0}^{d - 1} p_i \right)
    \ensuremath{\operatorname{mod}} (d \ell)$;
    
    \item $t' = \left( \ell a' + \sum_{i = 0}^{d - 1} p'_i \right)
    \ensuremath{\operatorname{mod}} (d \ell)$;
    
    \item $\delta = p'_a - p_a$; and
    
    \item $\delta' = p_{a'} - p'_{a'}$.
  \end{itemize}
  Thus, modulo $d \ell$,
  \begin{align*}
    t' - t + \delta' - \delta     \equiv &~ \ell (a' - a) + \sum_{i = 0}^{d - 1}
    (p'_i - p_i) - (p'_{a'} - p_{a'}) - (p'_a - p_a)\\
     \equiv &~ \ell (a' - a) + \sum_{\text{\scriptsize{$\begin{array}{c}
      i = 0\\
      i \not\in \{ a, a' \}
    \end{array}$}}}^{d - 1} (p'_i - p_i)\\
     \equiv &~ \ell (a' - a)
  \end{align*}
  since, for axes $i \not\in \{ a, a' \}$, we have $p_i = p'_i$ from the fact
  that the lines cross. However, there cannot be an intersection between $\bigl( (0 ; \ell) + t' - t +
  \delta' - \delta\bigr)
  \ensuremath{\operatorname{mod}} (d \ell)$ and $(0 ; \ell)
  \ensuremath{\operatorname{mod}} (d \ell) = (0 ; \ell)$ because $a \neq a'$.
  Therefore, {$(0 ; \ell) \cap \bigl( (0 ; \ell) + t' - t + \delta'
  - \delta\bigr) = \{ \}$} and there
  are no collisions.
\end{proof}

\begin{theorem}
  \label{thm:2d}Any regular 2-dimensional train network with trains of length
  $\ell$ has a schedule with delay at most $M - 1$, where
  \[ M = \left\{\begin{array}{ll}
       2, & \ell = 1\\
       8, & \ell = 2\\
       6 \ell, & \ell \geqslant 3.
     \end{array}\right. \]
\end{theorem}

\begin{proof}
  Assign delay
  \[ \sigma \Bigl( x + y 
       + (1 - a) \bigl( - 2 (y\ensuremath{\operatorname{mod}} \ell)
       - \ell + 1\bigr)
       + a \bigl( - 2 (x\ensuremath{\operatorname{mod}} \ell) + 2
       \ell - 1 \bigr)
     \Bigr) \ensuremath{\operatorname{mod}}M \]
  to lines departing from $(x, y)$ with axis $a \in \{ 0, 1 \}$ and sign
  $\sigma \in \{ - 1, 1 \}$. Consider then a horizontal line departing from
  $(x, y)$ with sign $\sigma$ that crosses a vertical line departing from
  $(x', y')$ with sign~$\sigma'$. In the notation of Definition
  \ref{def:schedule}, we have:
  \begin{itemize}
    \item $t = \sigma \bigl( x + y - 2
    (y\ensuremath{\operatorname{mod}} \ell) - \ell + 1\bigr)$;
    
    \item $t' = \sigma' \bigl( x' + y' - 2 (x'
    \ensuremath{\operatorname{mod}} \ell) + 2 \ell - 1\bigr)$;
    
    \item $\delta = \sigma (x' - x)$; and
    
    \item $\delta' = - \sigma' (y' - y)$.
  \end{itemize}
  Note that $M \geqslant 2 \ell$ and so it is enough to show that
  \[ (t' - t + \delta' - \delta) \ensuremath{\operatorname{mod}}M \in \{ \ell,
     \ldots, M - \ell \} . \]
  Indeed, then $(0 ; \ell) \ensuremath{\operatorname{mod}}M = (0 ; \ell)$ does
  not intersect $((0 ; \ell) + t' - t + \delta' - \delta)
  \ensuremath{\operatorname{mod}}M$, which implies~{$(0 ; \ell) \cap ((0 ;
  \ell) + t' - t + \delta' - \delta) = \{ \}$}, i.e., a lack of collisions. We
  prove this by cases starting when {$\sigma = \sigma' = 1$}, which gives us,
  modulo $M$,
  \begin{align*}
    t' - t + \delta' - \delta  \equiv &~ 
      x' + y' - 2 (x' \ensuremath{\operatorname{mod}} \ell) + 2
      \ell - 1  - y' + y
      - x - y + 2 (y\ensuremath{\operatorname{mod}} \ell) + \ell - 1 - x' + x
    \\
     \equiv &~ 2 (y\ensuremath{\operatorname{mod}} \ell) - 2 (x'
    \ensuremath{\operatorname{mod}} \ell)  + 3 \ell - 2.
  \end{align*}
  Because $(y\ensuremath{\operatorname{mod}} \ell), (x'
  \ensuremath{\operatorname{mod}} \ell) \in \{ 0, \ldots, \ell - 1 \}$, we
  have that
  \begin{align*}
  2 (y\ensuremath{\operatorname{mod}} \ell) - 2 (x'
     \ensuremath{\operatorname{mod}} \ell) + 3 \ell - 2 \in&~ \{ \ell, \ldots, 5
     \ell - 4 \}\\ \subseteq&~ \{ \ell, \ldots, M - \ell \} .  \end{align*}
  In case $\sigma = - 1$ and $\sigma' = 1$, we have, modulo $M$,
  \begin{align*}
    t' - t + \delta' - \delta  \equiv &~ 
      x' + y' - 2 (x' \ensuremath{\operatorname{mod}} \ell) + 2
      \ell - 1 - y' + y + x + y - 2 (y\ensuremath{\operatorname{mod}} \ell) - \ell + 1 + x' - x
    \\
    \equiv &~ 2 \bigl( x' - (x' \ensuremath{\operatorname{mod}}
    \ell) + y - (y\ensuremath{\operatorname{mod}} \ell)\bigr) + \ell .
  \end{align*}
  Because $x' - (x' \ensuremath{\operatorname{mod}} \ell)$ and $y -
  (y\ensuremath{\operatorname{mod}} \ell)$ are multiples of $\ell$, we must
  have {that $t' - t + \delta' - \delta$} is an odd multiple of $\ell$.
  Moreover, since $M$ is an even multiple of $\ell$, ${(t' - t +
  \delta' - \delta) \ensuremath{\operatorname{mod}}M}$ must indeed be in $\{ \ell, \ldots, M -
  \ell \}$.
  
  To complete the remaining two cases, note that if $\sigma$ and $\sigma'$
  were simultaneously multiplied by~$- 1$, so would be $t' - t + \delta' -
  \delta$. However, $\{ \ell, \ldots, M - \ell \}$ is closed under negation
  modulo $M$.
\end{proof}

\begin{theorem}\label{thm:3d}
  Any regular 3-dimensional train network with trains of length $1$ has a
  schedule with delay at most $5$.
\end{theorem}

\begin{proof}
  We delay each line by the only integer in $\{ 0, \ldots, 5 \}$ that is
  equivalent {to $\sigma (p_0 + p_1 + p_2 + a)$} {modulo $3$} and equivalent
  to $p_0 + p_1 + p_2 + (\sigma + 1) / 2$ modulo 2, where $p = (p_0, p_1, p_2)
  \in \mathbb{Z}^3$ is its departure point, $a \in \{ 0, 1, 2 \}$ is its axis
  and $\sigma \in \{ - 1, 1 \}$ is its sign (the existence and uniqueness of
  this integer is assured by the Chinese Remainder Theorem).
  
  Consider then two crossing lines that:
  \begin{itemize}
    \item Depart respectively from $p = (p_0, p_1, p_2) \in \mathbb{Z}^3$ and
    $p' = (p'_0, p'_1, p'_2) \in \mathbb{Z}^3$;
    
    \item Have respective signs $\sigma, \sigma' \in \{ - 1, 1 \}$; and
    
    \item Have $a \in \{ 0, 1, 2 \}$ and $a + 1 \in \{ 0, 1, 2 \}$ as their
    respective axes (axis arithmetic is done modulo~$3$).
  \end{itemize}
  Thus, in the notation of Definition \ref{def:schedule}, we have $\delta =
  \sigma (p'_a - p_a)$ {and $\delta' = - \sigma' (p'_{a + 1} - p_{a + 1})$}.
  Modulo~$2$, we also {have $t \equiv p_0 + p_1 + p_2 + (\sigma + 1) / 2$} and
  $t' \equiv p'_0 + p'_1 + p'_2 + (\sigma' + 1) / 2$. Therefore, in case~$\sigma \neq \sigma'$, also modulo $2$,
  \begin{align*}
    t' - t + \delta' - \delta  \equiv &~ 
      \sum_{i = 0}^2 (p'_i - p_i) + (\sigma' + 1) / 2 - (\sigma + 1) / 2 - \sigma' (p'_{a + 1} - p_{a + 1}) - \sigma (p'_a - p_a)
    \\
     \equiv &~ \sum_{i = 0}^2 (p'_i - p_i) - (p'_{a + 1} - p_{a + 1}) - (p'_a
    - p_a) + (\sigma' - \sigma) / 2\\
     \equiv &~ p'_{a + 2} - p_{a + 2} + (\sigma' - \sigma) / 2.
  \end{align*}
  However, since the lines cross, $p_{a + 2} = p'_{a + 2}$ and, since $\sigma,
  \sigma' \in \{ - 1, 1 \}$ and $\sigma \neq \sigma'$, it must be that
  {$(\sigma' - \sigma) / 2 \equiv 1$} modulo $2$, so $(t' - t + \delta' -
  \delta) \ensuremath{\operatorname{mod}}2 = 1$. The collision avoidance
  condition for these lines is {$(0 ; 1) \cap \bigl( (0 ; 1) + t' -
  t + \delta' - \delta\bigr) = \{ \}$}
  and must be true since $(0 ; 1) \ensuremath{\operatorname{mod}}2 = (0 ; 1)$
  but~$\bigl((0 ; 1) + t' - t + \delta' - \delta\bigr) \ensuremath{\operatorname{mod}}2
  = (1 ; 2)$.
  
  On the other hand, if $\sigma = \sigma'$, then note that, modulo $3$, 
  \[ t\equiv \sigma (p_0 + p_1 + p_2 + a) \quad\] and \[\quad {t' \equiv \sigma (p'_0 + p'_1 +
  p'_2 + a + 1)}. \] Therefore, again modulo $3$,
  \begin{align*}
    t' - t + \delta' - \delta  \equiv &~ 
      \sigma \left( \sum_{i = 0}^2 (p'_i - p_i) + a + 1 - a \right)- \sigma (p'_a - p_a) - \sigma (p'_{a + 1} - p_{a + 1})
    \\
     \equiv &~ \sigma (p'_{a + 2} - p_{a + 2} + 1)\\
     \equiv &~ \sigma
  \end{align*}
  as, once more, $p_{a + 2} = p'_{a + 2}$. As before, $(0 ; 1)
  \ensuremath{\operatorname{mod}}3 = (0 ; 1)$ but
  \[ \bigl( (0 ; 1) + t' - t + \delta' - \delta\bigr)
     \ensuremath{\operatorname{mod}}3 = (0 ; 1) + (\sigma
     \ensuremath{\operatorname{mod}}3), \]
  which is either $(1 ; 2)$ or $(2 ; 3)$. No collisions are therefore
  possible.
\end{proof}

\section{Clique Searching and Lower Bounds}\label{sec:clique}

Whether a train network admits an integer schedule with delay at most $D \in
\mathbb{N}$ can be decided with a clique search: create a graph $G$ containing
a vertex $v_{L, t}$ for each line $L$ and time $t \in \{ 0, \ldots, D \}$; for
every pair of distinct lines $L${ and $L'$} and for each pair of times $t, t'
\in \{ 0, \ldots, D \}$ which would not result in a collision if assigned as
delays respectively to $L$ and $L'$, put an edge {between $v_{L, t}$} {and~$v_{L', t'}$}. Note that if $L$ and $L'$ do not cross, then $v_{L, t} v_{L',
t'}$ is an edge for {all $t, t' \in \{ 0, \ldots, D \}$}. Note also that an
integer schedule with delay at most $D$ exists if, and only if, $G$ has a
clique with as many vertices as there are lines. This is because two vertices
associated with the same line cannot be selected and we encoded potential
collisions in the edges between vertices associated with different lines. So,
essentially, the clique is a delay assignment.

Due to Theorem \ref{thm:regular-integer-schedule}, we can decide whether a
regular train network has a schedule with delay at most a given number $D \in
\mathbb{R}$. We have implemented this algorithm with the Cliquer clique solver~{\cite{cliquer}} and recorded some lower bounds for regular train networks in
Table \ref{tab:lower-bounds}.

\begin{table}[h] 
	\centering
	\begin{tabular}{|c|c|c|c|c|c|}
		\hline
		$d$ & $\ell$ & $\sigma$ & Delay & Tight? & Figure\\
		\hline
		\hline
		2 & $\leq 10$ & $+$ & $2 \ell - 1$ & Theorem \ref{thm:positive-lines} &
		\ref{fig:lower-bounds}a\\
		\hline
		2 & 1 & $\pm$ & 1 & Theorem \ref{thm:2d} & \ref{fig:lower-bounds}a\\
		\hline
		2 & 2 & $\pm$ & 7 & Theorem \ref{thm:2d} & \ref{fig:lower-bounds}b\\
		\hline
		3 & 1 & $+$ & 2 & Theorem \ref{thm:positive-lines} &
		\ref{fig:lower-bounds}c\\
		\hline
	\end{tabular}
	
	\vskip0.5cm
	
	\caption{\label{tab:lower-bounds}A report on some useful regular train
		network classes including: the networks' dimension $d$, the length $\ell$ of
		their trains, whether their lines are all positive or of unrestricted sign,
		a lower bound on the delay of schedules for some networks in the class, our
		knowledge on whether this is an upper bound on the delay required to
		schedule all networks in the class and a reference to a figure describing a
		network that, when input to our algorithm, will produce the lower bound.}
\end{table}

\begin{figure}[h]\centering
	\includegraphics[page = 6]{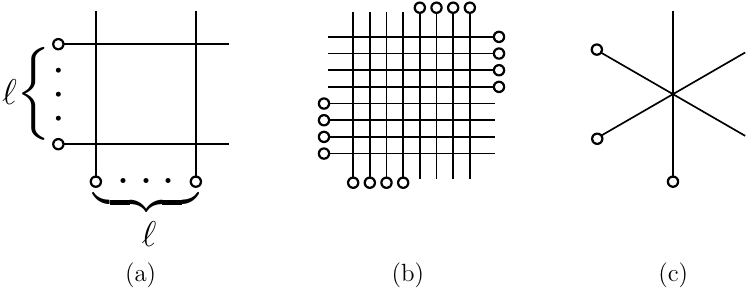}
	\caption{\label{fig:lower-bounds}Regular train networks ($\circ$ denotes the
		departure point of a line).}
\end{figure}

For greater exposition we will walk through a simple example of determining the lower bounds for a train network with only positive lines and a train length of $2$. Referring to Theorem \ref{thm:positive-lines} we see that a delay of $2\ell-1 = 4-1 = 3$ is possible. Therefore we will test a train network using the configuration in Figure \ref{fig:lower-bounds}a and a delay of $2$. If we construct a graph as outlined above and cannot find a clique of size $4$, then our configuration does not admit a network with a delay of $2$, and by Theorem \ref{thm:positive-lines} and Theorem \ref{thm:regular-integer-schedule} our lower bound on the delay must then be $3$. Although not necessary to our lower bound proof, for completeness we also show a graph built using a delay of $3$ and illustrate a clique of size $4$, which represents an assignment of delays that would result in no collisions.

We will express our train network in the following format:\footnote{This input format is a more readable version of the input we used to run our Python script, which built the graph in Cliquer readable format, and for which we refer the reader to the appendix.} \\
\texttt{<label> <train\_len> <axis><direction> <x> <y> <z>
}\\

\noindent where:\\

\begin{tabular}{l l}
\texttt{<label>} &is the line's label;\\
\texttt{<train\_len>}&	is the line's train length, \\
				& a positive integer;\\ 
\texttt{<axis>} & is the axis the track is\\
			&parallel to (``\texttt{x}", ``\texttt{y}", or ``\texttt{z}");\\
\texttt{<direction>} & is the line's direction of \\
				& movement (``-" or ``+");\\
\texttt{<x> <y> <z>} & are the line's departure\\ 
			& point.
\end{tabular}

\begin{figure}[h]
    \begin{subfigure}[b]{0.45\textwidth}\centering
	\includegraphics[width=5cm, page = 2]{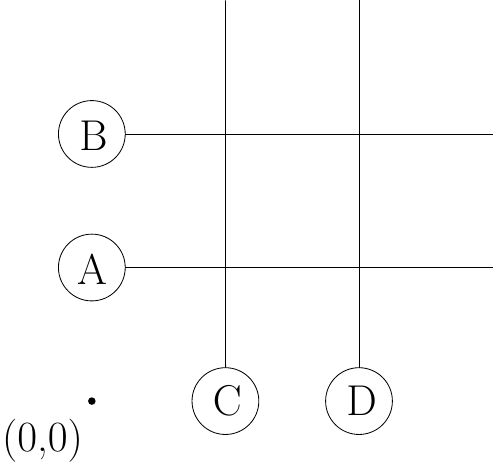}
	\caption{\label{fig:example1}Network 1. \textcolor{white}{ aaa aaa aaa aaa aaa\\ aaaa\\ aaa}}
	\end{subfigure}
    \begin{subfigure}[b]{0.45\textwidth}\centering
	\includegraphics[width=5.5cm, page = 6]{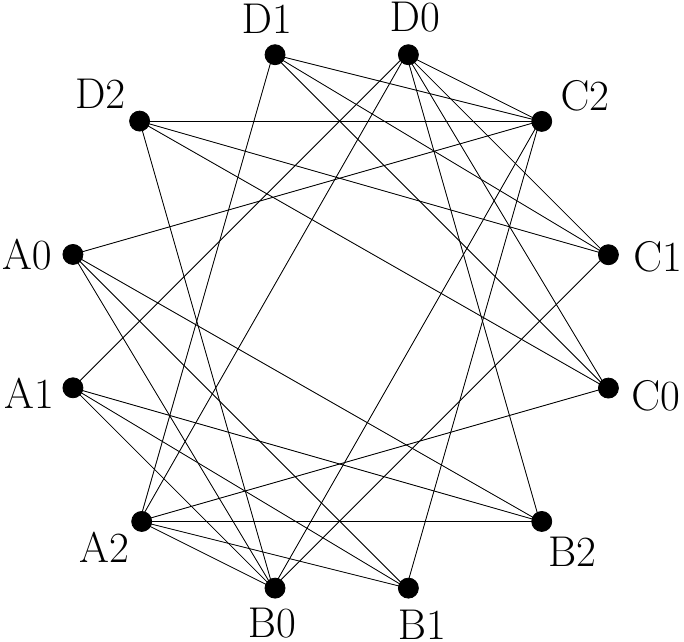}
	\caption{\label{fig:example2}The resulting graph for Network 1 and a maximum delay of $2$. The reader may verify there are no cliques of size $4$.}
	\end{subfigure}
	\caption{Network 1 and its associated graph with delay 2.}
\end{figure}

Our train network in the above format based on the configuration of Figure \ref{fig:lower-bounds}a and illustrated in Figure \ref{fig:example1}, is then:\\

\noindent A 2 x+ 0 1 0\\
B 2 x+ 0 2 0\\
C 2 y+ 1 0 0\\
D 2 y+ 2 0 0\\

\begin{figure}[h]\centering
	\includegraphics[width=7cm, page = 7]{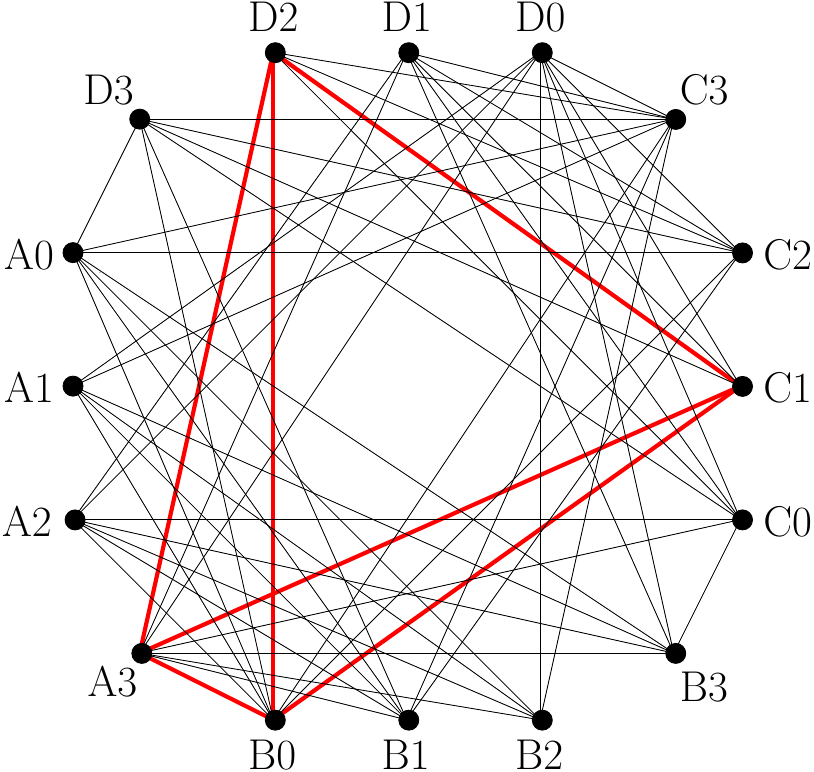}
	\caption{\label{fig:example3}The resulting graph for Network 1 and a maximum delay of 3. Note the clique of size $4$ which gives a delay assignment of A:3, B:0, C:1 and D:2.}
\end{figure}

We will refer to the above train network as Network 1. We create a vertex for each combination of line and possible delay, then for each pair of vertices consisting of distinct lines, we connect them with an edge if the delay assignment does not result in a collision. See Figure \ref{fig:example2}. We may then search this graph for cliques of size $4$, the number of train lines. To facilitate this search, we built our graph using a Python script, 
and fed the output into the Cliquer clique solver~\cite{cliquer}. Figure \ref{fig:example3} shows the graph resulting from Network 1 with a maximum delay of $3$, and the resulting clique which gives a delay assignment that will not result in any collisions.


\section{Open Problems}

The main problem that remains open is whether, for values of $\ell \geqslant
2$, every three-dimensional regular train network with trains of length $\ell$
admits a schedule with delay bounded by a constant. However, we are also
interested in extending the lower bounds in Table \ref{tab:lower-bounds} to
more values of $\ell$.

\section*{Acknowledgements}

We would like to thank Joseph S. B. Mitchell, Esther M. Arkin, Sándor Fekete
and members of the Computational Geometry group at Carleton University.

\small
\bibliographystyle{abbrv}

\begin{thebibliography}{99}

    \bibitem{AJAY2021}
	Ajay, Jammigumpula and Jana, Satyabrata and Roy, Sasanka
	\newblock Collision-free routing problem with restricted L-path.
	\newblock{\em Discrete Applied Mathematics},
	2021

	\bibitem{DM2015}
	Dasler, Philip and Mount, David M.
	\newblock On the complexity of an unregulated traffic crossing.
	\newblock {\em Algorithms and Data Structures},
	\newblock{Lecture Notes in Computer Science, vol. 9214, Springer},
	pages 224--235, 2015.
	
	
	\bibitem{HS2017}
	Hyla, P. and Szpytko, J.
	\newblock Automated guided vehicles: the survey.
	\newblock {\em Journal of KONES},
	24, 2017.
	
	
	\bibitem{ADK2016}
	Ajaykumar, Jammigumpula and Das, Avinandan and Saikia, Navaneeta and Karmakar, Arindam
	\newblock Problems on one way road networks.
	\newblock {\em Proceedings of the 28th Canadian Conference on Computational Geometry},
	pages 303--308, 2016.
	
	
	\bibitem{KT1991}
	Kim, Chang W. and Tanchoco, J. M. A.
	\newblock Conflict-free shortest-time bidirectional {AGV} routing.
	\newblock {\em The International Journal of Production Research},
	29(12):2377--2391, 1991.
	
	
	\bibitem{Koff1987}
	Koff, Gary A.
	\newblock Automatic guided vehicle systems: applications, controls and planning.
	\newblock {\em Material flow},
	4(1--2):3--16, 1987.
	
	
	\bibitem{ZWJ1991}
	Zeng, Laiguang and Wang, Hsu-Pin (Ben) and Jin, Song
	\newblock Conflict detection of automated guided vehicles: a {Petri} net approach.
	\newblock {\em The International Journal of Production Research},
	29(5):866--879, 1991.
	
	
	\bibitem{QHHW2002}
	Qiu, Ling and Hsu, Wen-Jing and Huang, Shell-Ying and Wang, Han
	\newblock Scheduling and routing algorithms for {AGVs:} a survey.
	\newblock {\em The International Journal of Production Research},
	40(3):745--760, 2002.
	
	
	\bibitem{Malmborg1990}
	Malmborg, Charles J.
	\newblock A model for the design of zone control automated guided vehicle systems.
	\newblock {\em The International Journal of Production Research},
	28(10):1741--1758, 1990.
	
	
	\bibitem{KTM1978}
	Kakikura, Masayoshi and Takeno, Jun Ichi and Mukaidono, Masao
	\newblock A tour optimization problem in a road network with one-way paths.
	\newblock {\em IEEJ Transactions on Electronics, Information and Systems},
	98(8):257--264, 1978.
	
	
	\bibitem{cliquer}
		Niskanen, Sampo and {\"O}sterg{\r a}rd, Patric R. J.
		\newblock Cliquer user's guide, version 1.0.
		\newblock {\em Communications Laboratory, Helsinki University of Technology, Espoo, Finland},
		Technical Report T48, 2003.
		
		
		\bibitem{Robbins1939}
		Robbins, Herbert Ellis
		\newblock A theorem on graphs, with an application to a problem of traffic control.
		\newblock {\em The American Mathematical Monthly},
		46(5):281--283, 1939.
		
		
		\bibitem{Scheffer2020}
		Scheffer, Christian
		\newblock Train scheduling hardness and algorithms.
		\newblock {\em The 13th International Conference and Workshop on Algorithms and Computation},
		\newblock{Lecture Notes in Computer Science, vol. 12049, Springer},
		pages 342--347,
		2020.
		
		
		\bibitem{ARM2000}
		Arora, Sudha and Raina, A. K. and Mittal, A. K.
		\newblock Collision avoidance among {AGVs} at junctions.
		\newblock {\em Intelligent Vehicles Symposium},
		2000.
		
		
		\bibitem{Vis2006}
		Vis, Iris F. A.
		\newblock Survey of research in the design and control of automated guided vehicle systems.
		\newblock {\em European Journal of Operational Research},
		170(3):677--709, 2006.
		
		
		\bibitem{YZQ2017}
		Yan, Xuejun and Zhang, Canrong and Qi, Mingyao
		\newblock {Multi-AGVs} collision-avoidance and deadlock-control for item-to-human automated warehouse.
		\newblock {\em Industrial Engineering, Management Science and Application (ICIMSA)},
		pages 1--5, 2017.
		
	
\end{thebibliography}

\onecolumn
\section*{Appendix}

We share our implementation, in Python 3, of a program to convert a train
network with a parameter $D \in \mathbb{N}$ into a graph that has a clique
with as many vertices as the network has lines if, and only if, there is an
integer schedule for the network with delay at most $D$. If the script were saved as, for example, trains.py, then a call to:\\

\noindent\texttt{./trains.py <D>}\\

\noindent where \texttt{<D>} is the parameter $D$ would initiate the script. Network 1 from Section \ref{sec:clique} would then be input from the command line as:\\

\noindent \texttt{2 x+ 0 1 0\\
2 x+ 0 2 0\\
2 y+ 1 0 0\\
2 y+ 2 0 0\\}

\noindent The output from this script is meant to be piped into (or read from a file)  the cliquer. We will give a brief overview of the format, but for more details we refer the reader to the Cliquer clique solver~\cite{cliquer} technical report. The first line of output would be:\\

\noindent \texttt{p graph <V> 0}\\

\noindent The character \texttt{p} indicates to the cliquer that this is the format line. The \texttt{graph} entry indicates the format, but it is there for consistency with older formats and is ignored by the cliquer. The \texttt{<V>} field is the number of vertices in the output graph. The \texttt{0} field indicates the number of edges, but the cliquer ignores this number as well, and determines the number of edges from the input. Each edge of the resulting graph would then be output on a separate line consisting of:\\

\noindent  \texttt{e <$v_1$> <$v_2$>}\\

\noindent where \texttt{$v_1$} and \texttt{$v_2$} are integers representing a vertex. The above output should be piped into the cliquer, which would output the largest clique found and the vertices contained therein. The line number (starting with line $0$) and delay assignment from a vertex $v$ on a network with $\ell$ lines total could be obtained with the following formulas:\\

\begin{tabular}{l l}
    Line number:  &  $(v-1)$ div $\ell$.\\
  Delay assignment: & $(v-1)$ mod $\ell$.\\
\end{tabular}\\

Our Python script, given below, operates from the command line but could be easily modified to read from and write to text files. 

\begin{verbatim}
#!/usr/bin/env python3

import sys
import itertools

# Standard input: one train line per input line, each in the form
#     <train_len> <axis><dir> <x> <y> <z>
# where:
#     * <train_len> is the line's train length, a positive integer;
#     * <axis> is the axis the track is parallel to ("x", "y", or "z");
#     * <dir> is the line's direction of movement ("-" or "+"); and
#     * <x>, <y> and <z> are the line's departure point.
#
# Command line argument: a non-negative integer D.
#
# Output: a graph in a Cliquer-compatible format that has a clique of size
#     equal to the number of train lines (or input lines) if, and only if, the
#     lines form a network that has an integer schedule with delay at most D.

class TrainLine:
    # train_len: a positive integer, the line's train length
    # start: a 3d point, the line's departure point
    # axis: the number of the axis the track is parallel to (0, 1 or 2)
    # dir: the line's direction of movement (-1 or 1)
    def __init__(l, train_len, start, axis, dir):
        l.train_len = train_len
        l.start = start
        l.axis = axis
        l.dir = dir

    # l[i] is the departure point's i-th coordinate (i is 0, 1 or 2)
    def __getitem__(l, i):
        return l.start[i]

    # Whether two line's tracks overlap
    def overlaps(l0, l1):
        if l0.axis != l1.axis:
            return False
        a = l0.axis
        if l0[(a+1)%3] != l1[(a+1)%3] or l0[(a+2)%3] != l1[(a+2)%3]:
            return False
        return l0.dir == l1.dir or l0[a]*l0.dir < l1[a]*l0.dir

    # Input: two train lines with non-overlapping tracks.
    # Output: None if the line's tracks do not cross and a pair with the
    # respective distances from the line's departure points to the
    # crossing point otherwise.
    def distances_to_crossing(l0, l1):
        other_axis = 0
        while other_axis == l0.axis or other_axis == l1.axis:
            other_axis+= 1
        if l0[other_axis] != l1[other_axis]:
            return None
        if l0[l1.axis]*l1.dir <= l1[l1.axis]*l1.dir:
            return None
        if l1[l0.axis]*l0.dir <= l0[l0.axis]*l0.dir:
            return None
        return (
            (l1[l0.axis]-l0[l0.axis])*l0.dir,
            (l0[l1.axis]-l1[l1.axis])*l1.dir
        )

    # Reads a train line from a text line
    @staticmethod
    def read(text_line):
        tokens = text_line.split()
        train_len = int(tokens[0])
        axis = {"x": 0, "y": 1, "z": 2}[tokens[1][0]]
        dir = {"-": -1, "+": 1}[tokens[1][1]]
        start = tuple(map(int, tokens[2:5]))
        return TrainLine(train_len, start, axis, dir)

# Whether two open intervals intersect
def intervals_overlap(I0, I1):
    (a0, b0) = I0
    (a1, b1) = I1
    return a0 < b1 and a1 < b0

# Command line argument
D = int(sys.argv[1])

# Read standard input into a list of train lines
L = [TrainLine.read(input_line) for input_line in sys.stdin.readlines()]

# Check input tracks are non-overlapping
for l0, l1 in itertools.combinations(L, 2):
    if l0.overlaps(l1):
        raise Exception("Input contains overlapping tracks")

# Assign vertices to lines
for i in range(len(L)):
    L[i].vertices = range((D+1)*i, (D+1)*(i+1))

# Print the graph
print("p graph %d 0" % ((D+1)*len(L)))
for l0, l1 in itertools.combinations(L, 2):
    ds = l0.distances_to_crossing(l1)
    if ds is not None:
        d0, d1 = ds
    for t0, t1 in itertools.product(range(D+1), repeat=2):
        if ds is None or not intervals_overlap(
            (t0 + d0, t0 + d0 + l0.train_len),
            (t1 + d1, t1 + d1 + l1.train_len)
        ):
            print("e %d %d" % (l0.vertices[t0]+1, l1.vertices[t1]+1))
\end{verbatim}

\end{document}